\documentclass[aps,prd,10pt,showpacs,groupedaddress,nofootinbib,letterpaper,twocolumn]{revtex4-1}

\usepackage{amsmath}
\usepackage{amsfonts}
\usepackage{amssymb}
\usepackage{mathrsfs}

\usepackage{graphicx}
\usepackage{color}
\usepackage{hyperref}
\usepackage{enumerate}

\begin{document}
   
\title{Unitarity of Maxwell theory on curved spacetimes in the covariant formalism}

\author{William Donnelly}
\affiliation{
Department of Applied Mathematics, \\
University of Waterloo \\
Waterloo, Ontario N2L 3G1, Canada}
\email{wdonnelly@uwaterloo.ca}
\author{Aron C. Wall}
\affiliation{
Department of Physics, \\
University of California, Santa Barbara \\
Santa Barbara, California 93106, USA
}
\email{aroncwall@gmail.com}

\begin{abstract}
Quantum field theory in curved spacetime may be defined either through a manifestly unitary canonical approach or via the manifestly covariant path integral formalism.
For gauge theories, these two approaches have produced conflicting results, leading to the question of whether the canonical approach is covariant, and whether the path integral approach is unitary.
We show the unitarity of the covariant U(1) Maxwell theory, defined via the Wick rotation of a Euclidean path integral.  
We begin by gauge-fixing the path integral, taking care with zero modes, large gauge transformations, and nontrivial bundles.  We find an extra geometric factor in the partition function that has been overlooked in previous work, coming from the zero mode of the gauge symmetry, which affects the entropy and stress-energy tensor.  
With this extra factor, the covariant calculation agrees with the canonical result for ultrastatic manifolds, and in $D = 2$.  Finally, we argue that if there exists a unitary (but not necessarily covariant) canonical formulation, then the covariant formulation must also be unitary, even if the two approaches disagree.
\end{abstract}


\maketitle

\section{Introduction}

The purpose of this article is to resolve some serious confusions about how to quantize gauge fields in curved spacetimes.  There are claims in the literature that the canonical and covariant methods of quantization lead to different results from each other \cite{Vassilevich1991, Vassilevich1992, Fukutaka1992a, Vassilevich1994}.
Analogous issues appear for gauge fields in the presence of dielectric media \cite{Bordag1997,Bei2011} and for linearized gravity in curved spacetime \cite{Fukutaka1992b}.
This is disturbing because it makes it unclear how to construct a theory that is simultaneously unitary and covariant.  

Since the Maxwell field is free, it is easy enough to promote the equations of motion to operator equations. 
But 
problems arise when one tries to calculate the dependence of the partition function $Z$ on the background geometry.  The geometry dependence is needed for e.g. calculations of the stress-energy tensor, or the geometric entropy \cite{Callan1994}.
To calculate the partition function one needs to gauge-fix the action, which can be done covariantly by introducing Faddeev-Popov ghosts.  
Although these ghosts do not interact with the gauge potential $A_{\mu}$, they do contribute to the geometry-dependent part of $Z$.  It is therefore important to check that they do not introduce any spurious unphysical effects.

We will focus on free Maxwell fields with gauge group U(1), on $D$-dimensional, spatially compact, connected\footnote{The assumption of connectedness simplifies notation, but is inessential since the partition function of a disconnected manifold is just the product of partition functions for each of its connected components.} and orientable spacetimes.  In section \ref{2} we will discuss the properties of the unfixed Maxwell partition function on compact Euclidean spacetimes without boundary.  In section \ref{3} we will derive the correct form of the covariant gauge-fixed theory in the continuum, including the correct normalization of the partition function $Z$, which is nontrivial.  This establishes the equivalence of the fixed and unfixed actions on compact manifolds.

For the most part, our analysis will recover the traditional ``vector minus two scalars'' Faddeev-Popov action.  But there are subtleties involving zero modes of the gauge symmetry and the gauge-fixing terms.  These subtleties lead to extra terms in the partition function which have not been properly taken into account in previous work.  

In section \ref{calc} we will consider two cases in particular where the partition function can be calculated by other methods: 1) the thermal partition function of pure electrodynamics on an ultrastatic manifold, and 2) the partition function on two-dimensional compact oriented manifolds.
We find that the above-mentioned extra terms must be included in the partition function in order to recover known results.

Finally, in section \ref{unitary} we will discuss Maxwell fields on nonstatic spacetimes with boundaries.  We will show that the unfixed covariant theory is unitary after Wick rotation to globally hyperbolic Lorentzian manifolds.  Our analysis shows that the covariant formulation of Maxwell fields is local and unitary, even if it turns out not to be equivalent to the canonical formulation.\footnote{We originally attempted to use BRST to prove unitarity, but it turns out that BRST is problematic for Maxwell fields on spatially compact manifolds.  The representation of BRST is reducible \cite{Barnich2000}, because there exists a canonically conjugate pair of spatially constant ghosts which are unpaired with any unphysical vector modes \cite{Donnelly2012}.  One can try to deal with this by adding ghosts-for-ghosts, but we were unable to deal with the resulting IR divergences in a satisfactory way.}

\section{A Local Covariant Action}\label{2}

We start with the Euclidean Maxwell partition function, prior to gauge-fixing:
\begin{equation}\label{unfix}
Z = \sum_\mathrm{bundles} \int \frac{DA}{\mathrm{Vol}(G)} e^{-S[F]},
\end{equation}
where $S$ is the standard Maxwell action given by
\begin{equation}
S = \int d^Dx\,\sqrt{g}\,F_{\mu\nu}F^{\mu\nu}/4,
\end{equation}
$\mathrm{Vol}(G)$ is the volume of the gauge group $G$ consisting of all local gauge transformations defined by $\delta A = \nabla \alpha$ where $\alpha$ is a scalar parameter which is periodic in the range $[0,\,2\pi/q)$, $q$ being the fundamental unit of charge associated with the U(1) gauge symmetry.
When the value of $\alpha$ increases by a nonzero multiple of $2 \pi / q$ going around a closed curve it represents a large gauge transformation; otherwise it is a small gauge transformation.

Because the gauge group is U(1), the integral of $A$ around around a closed curve is defined only modulo $2\pi /q$.
As a result there can exist nontrivial bundles corresponding to magnetic instantons.
To sum over these nontrivial bundles, one simply sums up $e^{-S}$ for all possible harmonic 2-forms satisfying the Dirac quantization conditions.

The path integral measure $DA$ is defined as follows:
If one writes out modes of $A$ in an orthonormal basis:
\begin{equation}
A_\mu = \sum_n A_n v^n_\mu, \qquad \int d^Dx \sqrt{g} v^m_\mu v^n_\nu g^{\mu \nu} = \delta_{mn},
\end{equation}
then the path integral measure can be written out explicitly as
\begin{equation}\label{Amu}
DA = \prod_n \mu_A \,dA_n,
\end{equation}
where $\mu_A$ is a factor with dimensions of mass required to keep $Z$ dimensionless.  Similarly, in order to define ${\mathrm{Vol}(G)}$, one needs a measure on the space of gauge transformations:
\begin{equation}\label{alphamu}
D\alpha = \prod_n \mu_\alpha^2 \,d\alpha_n,
\end{equation}
where $\mu_\alpha^2$ is a factor with mass-dimension two.\footnote{These factors of $\mu$ are similar to a UV cutoff $\Lambda$, insofar as they are dimensionful parameters needed to make the path integral well-defined.  Indeed, in some regulator schemes (such as the heat kernel), the same dimensionful parameter plays the role of both $\Lambda$ and $\mu$.  Nevertheless, it is important to recognize that this $\mu$ does not result from UV divergences, and can therefore play a role even for $D = 2$ Maxwell fields, where there are no local degrees of freedom.}

The partition function (\ref{unfix}) has the advantage of being manifestly covariant.  It is also manifestly local, in the sense that $e^{-S}$ and the measure are products of local functions of the field variables.  Unfortunately, it is also slightly ill-defined, because the pure-gauge modes of $A$ have infinite fluctuations, while $\mathrm{Vol(G)}$ is correspondingly infinite.  Formally, these two infinities cancel each other out to get a finite answer: the precise finite value of $\infty/\infty$ can be fixed by cutting off each pure gauge mode of $A$ at the same value as the corresponding gauge symmetry mode $\nabla \alpha$.  The zero mode $\alpha_0$ corresponding to constant gauge transformations needs no cutoff, because it is periodic due to the U(1) nature of the gauge symmetry.

The reader may wonder whether it is really correct to include the zero mode $\alpha_0$, since this transformation leaves the $A$ field unchanged.  The answer is yes.  In its current form, $e^{-S}$ and the measure are local functions of the fields, being a product of contributions associated with each point of space.  But if $\alpha_0$ were dropped, this would no longer be the case.  $\mathrm{Vol}(G)$ would be a nonlocal function of the geometry.    In section \ref{unitary} we will see that this locality property is needed to derive the unitarity of the theory.  Hence, just as Feynman diagrams which are invariant under a permutation group $Q$ have to have their amplitude multiplied by a symmetry factor $1/|Q|$, so also a spacetime history whose fields are invariant under a global gauge symmetry must have its amplitude divided by the volume of the $\alpha_0$ moduli space.  Integrating using the measure (\ref{alphamu}), one finds that the path integral has an overall symmetry factor
\begin{equation}
\frac{1}{{\mathrm{Vol}(\alpha_0)}} = \frac{q}{2\pi \mu_\alpha^2 \sqrt{V}},
\end{equation}
where $V$ is the volume of spacetime.  This factor affects the trace of the stress-energy tensor
\begin{equation}
T = \frac{\delta \ln Z}{\delta \sqrt{g}}
\end{equation} 
as well as the entropy of a thermal manifold with inverse temperature $\beta$:
\begin{equation}
S = (1 - \beta \frac{\partial}{\partial \beta}) \ln Z
\end{equation}
In section \ref{calc} we will show that this factor is necessary in order to obtain correct results for thermodynamic calculations.

\section{Gauge-Fixing}\label{3}

We now perform the Feynman-Faddeev-Popov-t'Hooft trick in order to gauge-fix the small gauge transformations.  We would like to insert into the path integral a factor of unity which includes the delta function
\begin{equation}
\delta(\nabla_\mu A^\mu - \omega),
\end{equation}
with $\omega$ being an auxilliary scalar parameter.
However this is overconstraining, since on a compact surface one of the constraints is redundant (if the integral of $\omega$ is zero) or impossible (if the integral of $\omega$ is nonzero).

To deal with this in a covariant way, we use a complete basis of orthonormal scalar modes $\phi_n$ of the Laplacian:
\begin{equation}
-\nabla^2 \phi_n = \lambda_n \phi_n, \qquad \int d^Dx \sqrt{g} \phi_n^2 = 1.
\end{equation}
This allows us to expand out $\omega$ and $\nabla_\mu A^\mu$ in orthonormal modes of the Laplacian:
\begin{equation}
\omega = \sum_n \omega_n \phi_n, \qquad \nabla_\mu A^\mu = \sum_n (\nabla_\mu A^\mu)_n \phi_n.
\end{equation}
We can then insert into the path integral a factor of unity which includes every component of the delta function \emph{except} the zero mode:
\begin{equation}
1 = \int D'\alpha \prod_{n \ne 0} \delta((\nabla_\mu A^\mu_{(\alpha)})_n - \omega_n) \mathrm{det}'(-\mu_\alpha^{-2} \nabla^2),
\end{equation}
where $A^\mu_{(\alpha)} = A^\mu + \nabla^\mu \alpha$,
$\mathrm{det}'(-\mu_\alpha^{-2} \nabla^2)$ is the usual Faddeev-Popov determinant, and the primes are reminders to omit zero modes.
In the case of $D'\alpha$ we also omit the large gauge transformations, which will be handled separately.

The determinant can be viewed as the quantum field theory of two scalar Grassmannian fields known as Faddeev-Popov ghosts, whose partition function is given by
\begin{equation}
\mathrm{det}'(-\mu_\alpha^{-2}\nabla^2) = \int D'c\,D'\bar{c}\,e^{-\int d^Dx\,\bar{c}\,\nabla^2 c},
\end{equation}
where the Grassmannian integrals are defined by
\begin{equation} \label{Dc}
\int D'c\,D'\bar{c} = \prod_{n \ne 0} \mu_\alpha^{-2} \frac{\partial}{\partial c_n} \frac{\partial}{\partial\bar{c}_n} 
\end{equation}
with the zero modes of $c$ and $\bar{c}$ omitted.
The factor of $\mu$ in Eq.~\eqref{Dc} is necessary to make the partition function dimensionless; the fact that it is the same $\mu_\alpha$ that appears in Eq.~\eqref{alphamu} is required so that the Faddeev-Popov trick is equivalent to inserting an identity operator.

One can then perform the $D\alpha$ integrals by changing variables from $A$ to $A_{(\alpha)}$, making the integrand independent of $\alpha$, leading to a factor of $\mathrm{Vol}(G)$ that cancels (formally) with the one in Eq.~\eqref{unfix}.
This eliminates all gauge symmetry except for the zero mode and large gauge transformations, giving the following gauge-fixed partition function:
\begin{widetext}
\begin{equation}
Z = \frac{q}{2\pi \mu_\alpha^2 \sqrt{V}} \sum_\mathrm{bundles} 
\int 
\frac{DA}{G_\mathrm{large}}
 D'c\,D'\bar{c}\,
\prod_{n \ne 0} \delta((\nabla_\mu A^\mu)_n - \omega_n)\
e^{-S[F, c, \bar{c}]}.
   \end{equation}
The measure $DA / G_\mathrm{large}$ is an integration over equivalence classes of $A$ under large gauge transformations; formally it can be thought of as dividing by the number of Gribov copies, i.e. configurations of $A$ which satisfy $\nabla_\mu A^\mu = 0$ and are equivalent to $A = 0$ under a large U(1) gauge-transformation.
\end{widetext}

We can then give dynamics to the unphysical modes of $A$ by integrating over $\omega$ using the identity
\begin{equation} \label{N}
\frac{1}{N(\xi)} \int D\omega \,e^{- \int d^Dx \sqrt{g} \omega^2/2\xi} = 1,
\end{equation}
where $N$ is an infinite normalization factor.  Since this normalization factor is local, and depends only on the spacetime volume $\sqrt{g}$, it can be dropped by absorbing it into a redefinition of the cosmological constant.  One can then perform the $D\omega$ integrals to obtain the following partition function:
\begin{equation}\label{fixed}
Z = \frac{q}{\mu_\alpha^2} \sqrt{\frac{\xi}{2\pi V}}
\sum_\mathrm{bundles} \int 
\frac{DA}{G_\mathrm{large}}
D'c\,D'\bar{c}\,
e^{-S[A, c, \bar{c}]},
\end{equation}
where the action is the usual covariant action in t'Hooft $R_\xi$ gauge
\begin{equation}
S = \int d^Dx\,\sqrt{g}\,\frac{1}{4}F_{\mu\nu}F^{\mu\nu} + \frac{1}{2\xi} (\nabla_\mu A^\mu)^2 - \bar{c}\,\nabla^2 c,
\end{equation}
and there is an extra factor coming from the Gaussian integral over the unmatched zero mode $\omega_0$:
\begin{equation}
\int d\omega_0\,e^{-\omega_0^2/2\xi} = \sqrt{2\pi \xi}.
\end{equation}
In this gauge-fixed form of the partition function, covariance is manifest, but locality has been hidden because the zero modes of the field are treated on a different footing than the nonzero modes.  Nevertheless, the theory is equivalent by construction to the original unfixed action (\ref{unfix}).

To further evaluate the partition function of Eq.~\eqref{fixed}, we divide the fields $A_\mu$ into the zero modes and non-zero modes.
The non-zero modes can be evaluated directly as Gaussian integrals,
\begin{align}
Z_T &= \prod_{\lambda \in \sigma_T} \mu_A \sqrt\frac{2 \pi}{\lambda} \\
Z_L &= \prod_{\lambda \in \sigma_S} \mu_A \sqrt\frac{2 \pi \xi}{\lambda} \\
Z_G &= \prod_{\lambda \in \sigma_S} \mu_\alpha^{-2} \lambda
\end{align}
Here $Z_T$ is the path integral over transverse modes, which depends on $\sigma_T$, the nonzero spectrum of the Hodge Laplacian on $1$-forms restricted to act on transverse modes.  
$Z_L$ is the path integral over longitudinal modes, and $Z_G$ is the path integral over ghosts; both of these depend on $\sigma_S$, the nonzero spectrum of the scalar Laplacian.

Now consider the zero modes of $A$.
The path integral over zero modes does not have the form of a Gaussian integral, as was the case for the non-zero modes.
The path integral over these modes would be infinite, were it not cut off by identifying configurations related by large gauge transformations.
Large gauge transformations identify $A$ and $A + (2 \pi/ q) \eta$ for $\eta \in H^1(M, \mathbb{Z})$, where the cohomology group $H^1(M, \mathbb{Z})$ is the set of harmonic $1$-forms whose holonomies are all integer-valued.
The space of flat connections modulo large gauge transformations is the moduli space $\mathscr{M}$, and the path integral over zero modes is equal to its volume $\text{Vol}(\mathscr{M})$.

It will be convenient to adopt a non-orthonormal basis for $\mathscr{M}$ consisting of harmonic $1$-form fields $w_1, \ldots, w_{b_1}$ that form a $\mathbb{Z}$-basis for $H^1(M,\mathbb{Z})$, i.e. such that every $1$-form with integer-valued holonomies can be written as a linear combination of the $w_I$ with integer coefficients.
In this basis, the flat connections modulo gauge transformations is the region $[0,2 \pi / q)^{b_1}$. 
The price to pay for this choice of basis is that the path integral measure acquires a factor $\sqrt{\det(\Gamma)}$, where $\Gamma$ is the metric on moduli space:
\begin{equation} \label{Gamma}
\Gamma_{IJ} = \int w_I \wedge \star w_J = \int d^Dx \sqrt{g} \, g^{ab} w_{Ia} w_{Jb}.
\end{equation}
The volume of the moduli space is therefore
\begin{equation}
\text{Vol}(\mathscr{M}) = \left( \frac{2 \pi \mu_A}{q} \right)^{b_1} \sqrt{\det(\Gamma)}.
\end{equation}

Finally, we have the monopole configurations.
These are indexed by harmonic two-form fields $F$ satisfying the Dirac quantization condition, $F \in \frac{2 \pi}{q} H^2(M, \mathbb{Z})$, where the cohomology group $H^2(M, \mathbb{Z})$ is the set of 2-forms having integer flux through every closed two-dimensional surface $S \subset \Sigma$.
The contribution of these field configuration to the partition function is
\begin{equation}
Z_\text{bundles} = \sum_{F \in \frac{2 \pi}{q} H^2(M, \mathbb{Z})} e^{-S[F]}
\end{equation}

Thus combining all these factors we obtain
\begin{equation} \label{Zfinal}
Z = \frac{q}{\mu_\alpha^2} \sqrt{\frac{\xi}{2\pi V}}
\text{Vol}(\mathscr{M})
Z_T Z_L Z_G
\sum_{F \in \frac{2 \pi}{q} H^2(M, \mathbb{Z})} e^{-S[F]}
\end{equation}

\section{Examples}\label{calc}

\subsection{Two-dimensional case}

Let us consider the case where the manifold $M$ is two-dimensional and orientable.
In two dimensions, the Maxwell field has no local degrees of freedom.
This is reflected in the fact that the path integral simplifies considerably, due to a cancellation between the gauge potential and ghosts.
In particular, the nonzero spectrum $\sigma_T$ of the transverse Laplacian is identical to the nonzero spectrum of the scalar Laplacian $\sigma_S$, so the contribution of the nonzero modes is
\begin{equation}
Z_T Z_L Z_G = \prod_{\lambda \in \sigma_S} \frac{\mu_A^2}{\mu_\alpha^2} 2 \pi \sqrt{\xi}
= \frac{\mu_\alpha^2}{\mu_A^2} \frac{1}{2 \pi \sqrt{\xi}}
\end{equation}
Here we used the fact that if the product had been a product over all scalar modes (not just non-zero modes) then it could be absorbed into a local term.\footnote{In this step we are essentially just reintroducing the factor of $N(\xi)$ that was dropped from Eq.~\eqref{N}.}
Note that this factor of $\sqrt{\xi}$ cancels the one in Eq.~\eqref{fixed}, as required by gauge-invariance of the partition function.

To compute the volume of the moduli space, we appeal to Poincar\'e duality to show that the metric $\Gamma$ \eqref{Gamma} on moduli space has unit determinant.\footnote{For a concise summary of the relevant aspects of Poincar\'e duality see Ref.~\cite{Headrick2012}}
In two dimensions, Poincar\'e duality implies that under the bilinear pairing
\begin{equation} \label{pairing}
\langle \eta, \zeta \rangle = \int \eta \wedge \zeta,
\end{equation}
the lattice $H^1(M, \mathbb{Z})$ is self-dual:
\begin{equation}
\{ \eta : \langle \eta, \zeta \rangle \in \mathbb{Z} \quad \forall \zeta \in H^1(M, \mathbb{Z}) \} = H^1(M, \mathbb{Z}).
\end{equation}
Expressed in the basis $w_I$, this implies that the matrix
\begin{equation}
P_{IJ} = \int w_I \wedge w_J
\end{equation}
is invertible over the integers, and therefore $\det(P) = \pm 1$.
The metric $\Gamma_{IJ}$ is related to the bilinear form \eqref{pairing} by
\begin{equation}
\Gamma_{IJ} = \left \langle w_I, \star w_J \right \rangle
\end{equation}
where $\star$ denotes the Hodge star.
In terms of matrices, $\Gamma = P S$ where $S$ is the matrix $\star w_I = S_{IJ} w_J$.
Since $\star^2 = -1$, the matrix $S$ satisfies $\det(S) = \pm 1$.
The determinant of $\Gamma$ is then $\det \Gamma = \det P \det S = 1$ (the positive sign follows from the fact that $\Gamma$ is positive-definite).
Thus in two dimensions, the volume of the moduli space is
\begin{equation}
\text{Vol}(\mathscr{M}) = \left( \frac{2 \pi \mu_A}{q} \right)^{b_1} \sqrt{\det{\Gamma}} = \left( \frac{2 \pi \mu_A}{q} \right)^{b_1}.
\end{equation}

On a two-dimensional connected manifold, the only harmonic two-forms are proportional to the volume form $\epsilon_{ab}$, and the Dirac quantization condition gives
\begin{equation}
F_{ab} = \frac{2 \pi}{q} \frac{n}{V} \epsilon_{ab}
\end{equation}
where $n \in \mathbb{Z}$.
Their contribution to the partition function is
\begin{equation}
Z_\text{bundles} = \sum_{n \in \mathbb{Z}} e^{-2 \pi^2 n^2 / V q^2}.
\end{equation}
Combining these factors we obtain
\begin{equation}
Z = \sqrt\frac{2 \pi}{q^2 V} \left( \frac{2 \pi \mu_A}{q} \right)^{b_1 - 2} \sum_{n \in \mathbb{Z}} e^{-2 \pi^2 n^2/V q^2}.
\end{equation}
Since we assume $M$ is connected and orientable, the exponent $b_1 - 2 = -\chi$, the Euler characteristic of $M$. 
Using the Gauss-Bonnet theorem $\chi = \frac{1}{4 \pi} \int \sqrt{g} R$, this term can be absorbed into a local counterterm (the Einstein-Hilbert term) and we therefore discard it.
By Poisson summation the remaining part of the partition function can be put in the canonical form
\begin{equation}
Z = \sum_{n} e^{-\frac{1}{2} V q^2 n^2}.
\end{equation}
This agrees with the known result for the partition function of two-dimensional gauge theory, calculated using topological quantum field theory methods \cite{Cordes1994}.

\subsection{Ultrastatic manifold}

We now consider the ultrastatic case in which $M$ is of the form $S^1 \times \Sigma$, where the $S^1$ direction is the imaginary time coordinate $\tau$ with period $\beta$ and $\Sigma$ is compact and orientable, but otherwise arbitrary.
We will show that the formula \eqref{Zfinal} reproduces the canonical form of the partition function $Z = \text{tr}\,e^{-\beta H}$.  We divide the field modes as follows:
\begin{enumerate}
\item Nonharmonic modes of $\Sigma$:
	\begin{enumerate}
	\item transverse spatial polarizations; \label{tran}
	\item longitudinal spatial polarizations; \label{long}
	\item temporal polarizations; \label{temp}
	\item ghosts; \label{ghos}
	\end{enumerate}
\item Harmonic modes of $\Sigma$:
\begin{enumerate}
\setcounter{enumii}{4}
	\item vector zero modes of $\Sigma$; \label{v0}
	\item spatially constant temporal modes; \label{t0}
	\item spatially constant ghosts; \label{g0}
	\end{enumerate}
\item Nontrivial U(1) bundles on $\Sigma \times S^1$. \label{bund}
\end{enumerate}
In the first category, there is a factor of $(2 \sinh (\beta \sqrt{\lambda}/2))^{-1}$, the canonical partition function of a mode with frequency $\sqrt{\lambda}$, for each transverse vector mode with Laplacian eigenvalue $\lambda$.  The contributions of the remaining modes to $Z$ cancel each other exactly.

\begin{widetext}
The second category includes, not only the spatially constant temporal mode and ghosts, but also $b_1 - 1$ vector zero modes of $\Sigma$.  Here $b_1$ is the Betti number of $M$ (which counts the number of harmonic vectors); we subtract 1 from this because the temporal mode is included separately.  Excluding the integral over vectors that are constant in time (the moduli space), the contribution of all these modes to $Z$ is
\begin{equation}
\prod_{n \neq 0} \mu_\alpha^{-2}  \left( \frac{2 \pi n}{\beta} \right)^2 \mu_A \sqrt{2 \pi \xi}  \left( \frac{2 \pi n}{\beta} \right)^{-1} \left[ \mu_A \sqrt{2 \pi}  \left( \frac{2 \pi n}{\beta} \right)^{-1} \right] ^{b_1-1} \\
= \frac{\mu_\alpha^2 \beta^2}{\sqrt{\xi}} \left( \frac{1}{\sqrt{2 \pi} \mu_A \beta} \right)^{b_1}, \label{spaceconstant}
\end{equation}
where we have used $\zeta$-function regularization of the product, $\prod_{n \geq 1} a n = a^{\zeta(0)} e^{- \zeta'(0)} = \sqrt{2 \pi / a}$. 
Note that the factor $\xi^{-1/2}$ cancels the explicit factor of $\xi^{1/2}$ in Eq.~\eqref{Zfinal}, as required by gauge invariance.
\end{widetext}

Next we calculate the volume of the moduli space.
A basis for the cohomology group of $M$ is given by a one-form $w_1 = \beta \, d \tau$ wrapping the time direction together with a basis $w_2, \ldots, w_{b_1}$ of 
$H^1(\Sigma, \mathbb{Z})$.
Letting $\gamma$ denote the metric on the moduli space of $\Sigma$, the metric $\Gamma$ on the moduli space of $M$ is given by
\begin{equation}
\Gamma = 
\left[ 
\begin{array}{cc}
\frac{V_\Sigma}{\beta} & 0 \\ 
0 & \beta \gamma_{IJ}
\end{array}
\right]
\end{equation}
and hence its volume is
\begin{equation}
\text{Vol}(\mathscr{M}) = \left( \frac{2 \pi \mu_A}{q} \right)^{b_1} \beta^{(b_1 - 2)/2} \sqrt{V_\Sigma \, \det{\gamma}} \label{volMstatic}.
\end{equation}

The final kind of contribution comes from nontrivial U(1) bundles.  These are proportional to suitably quantized harmonic 2-forms on $M$.  They can be divided into two types: harmonic 2-forms on $\Sigma$ constant in $\tau$ (magnetic fields), and those of the form $d\tau \wedge w_I$ where $w_I$ is a harmonic 1-form on $\Sigma$ (electric fields).

The magnetic fields are spatial 2-forms $B$ satisfying the Dirac quantization condition $\int_{S} B \in \frac{2 \pi}{q} \mathbb{Z}$ for every closed two-dimensional surface $S$.
These contribute to the partition function as
\begin{equation}
Z_B = \sum_{B} e^{- \frac{\beta}{2} \int_\Sigma d^{D-1} x \sqrt{q} B^2}
\end{equation}
which is precisely the contribution of the magnetic monopoles to the canonical partition function.

Now consider the electric fields: 2-forms proportional to $d \tau \wedge w_I$.
If we integrate over a surface $S^1 \times \gamma_I$ the quantization condition implies that $F = \frac{2 \pi}{q \beta} n_I d \tau \wedge w_I$ where $n_I \in \mathbb{Z}$.
The contribution of these field configurations to the partition function is
\begin{equation} \label{ZE}
Z_E = \sum_{n \in \mathbb{Z}^{b_1-1}} \exp \left[-\frac{1}{2 \beta} \left( \frac{2 \pi}{q} \right)^2 \sum_{I,J} \gamma_{IJ} n_I n_J \right].
\end{equation}
This is not of the canonical form, because the factor of $\beta$ in the exponent of Eq.~\eqref{ZE} appears in the denominator.
However, we can remedy this using the Poisson summation formula,
\begin{equation}
\sum_{x \in \mathbb{Z}^n} e^{-x^T M x} = \det \left(\frac{M}{\pi} \right)^{-1/2} \sum_{y \in \mathbb{Z}^n} e^{-\pi^2 y^T M^{-1} y},
\end{equation}
where $M = \frac{1}{2 \beta} \big( \frac{2 \pi}{q} \big)^2 \gamma$.
Thus we can rewrite Eq.~\eqref{ZE} as 
\begin{widetext}
\begin{equation} \label{ZEpoisson}
Z_E = \left( \frac{q^2 \beta}{2 \pi} \right)^{(b_1 - 1)/2} \frac{1}{\sqrt{\det(\gamma)}} \sum_{m \in \mathbb{Z}^{b_1 - 1}} \exp \left[ - \frac{\beta \, q^2}{2} \gamma^{IJ} m_I m_J \right].
\end{equation}
where $\gamma^{IJ}$ is the inverse of $\gamma_{IJ}$ and Einstein summation is assumed.
\end{widetext}

The prefactors in Eqs.~\eqref{Zfinal},\eqref{spaceconstant},\eqref{volMstatic},\eqref{ZEpoisson} cancel, so the path integral has the form of a canonical partition function.
All that remains to show is that the sum in Eq.~\eqref{ZEpoisson} is the same as the canonical partition function of the harmonic $E$-fields satisfying the quantization condition $\int_S E \in q \mathbb{Z}$ for all closed $(D-2)$-surfaces $S$.
In what follows we treat $E$ as an $(D-2)$-form related to the usual vector field $E$ by Hodge duality of vector fields on $\Sigma$.
By Poincar\'e duality, the condition that $\int_S E \in q \mathbb{Z}$ for closed $S$ is equivalent to $\int w_I \wedge E = q m_I$ where $m_I \in \mathbb{Z}$ for all $I$.
Since $E$ is a harmonic ($D-2$)-form, it can be expanded in the basis $\{ \star w_I \}$, and takes the form
\begin{equation}
E = q \gamma^{IJ} m_I \star w_J.
\end{equation}
The energy of such an $E$-field is
\begin{equation}
\frac{1}{2} \int E \wedge \star E = \frac{q^2}{2} \gamma^{IJ} m_I m_J.
\end{equation}
Thus the sum in Eq~\eqref{ZEpoisson} is precisely the canonical partition function of the harmonic $E$-fields satisfying the quantization condition.

The path integral expression \eqref{Zfinal} therefore gives the same result as the canonical partition function.  Hence the covariant formulation of quantum electrodynamics is unitary on ultrastatic manifolds.  

Note that this result depends crucially on using the correct normalization factor in the gauge-fixed partition function \ref{Zfinal}.  Without this prefactor, we would not get the right result even in the trivial case of $D = 1$ electromagnetism, where (because $\Sigma$ is 0-dimensional) there are no physical degrees of freedom at all.  Yet the Faddeev-Popov action has two ghosts but only one vector mode, so it appears at first sight that there are -1 scalar degrees of freedom.
This spurious contribution is cancelled when the zero modes are properly handled.

\section{Non-Static Unitarity}\label{unitary}

On non-static manifolds, it is not clear that the covariant and canonical approaches agree with one another \cite{Vassilevich1994}.  Nevertheless, we will argue that the covariant partition function must be unitary once it is Wick rotated to Lorentzian signature (assuming without proof that this Wick rotation is possible).   Our argument will rely on the \emph{existence} of a unitary canonical approach to electromagnetism, but we will not assume that the two approaches agree in all respects.

In this section, we will go back to the original unfixed form (\ref{unfix}) of the covariant Euclidean partition function $Z$.  In order to define unitarity, we need to be able to evaluate $Z$ on a manifold $M$ with boundary $\partial M$.

\subsection{Manifolds with boundary}

\medskip\textbf{States.}
The partition function of a manifold $M$ with boundary $\partial M$ is a functional of the geometry of $M$ and of a specified boundary value of the dynamical fields.
Because the unfixed action does not propagate the temporal or longitudinal modes, in order to compute $Z(M)$ it is sufficient to specify the connection on $\partial M$.  Let $A_{\partial M}$ refer to the pullback of the vector potential to $\partial M$ (which includes only the $D-1$ components of $A$ which point along $\partial M$).  Then the connection is specified by $A_{\partial M}$ modulo gauge transformations $\alpha_{\partial M}$.

Since the value of $Z(M)$ depends on $A_{\partial M}$, the path integral can be regarded as outputting a wavefunction $\Psi[A_{\partial M}]$ on the space of boundary vector potentials.  This wavefunction can be written as
\begin{equation}\label{state}
\Psi[A_{\partial M}] = \sum_\mathrm{bundles} \int_{A_{\partial M}} \frac{DA_M}{\mathrm{Vol}(G_M)} e^{-S_M[F]},
\end{equation}
where the integral is over connections with a specified pullback to $\partial M$ and 
$G_M$ refers to the group of gauge transformations $\alpha$ which vanish on $\partial M$.  However, gauge symmetry implies that $\Psi$ is constant along modes of $A$ which are pure gauge on $\partial M$.

\medskip\textbf{Inner Product.}  For any region $\Sigma \subset \partial M$, the inner product on the space of states is given by
\begin{equation}\label{inner}
\langle \Phi, \Psi \rangle_\Sigma = \sum_\mathrm{bundles} \int \frac{DA_\Sigma}{\mathrm{Vol}(G_\Sigma)} \Phi^*[A_\Sigma] \Psi[A_\Sigma],
\end{equation}
where $G_\Sigma$ refers to gauge transformations on $\Sigma$ that vanish on $\partial \Sigma$.  Any two open manifolds $M_1$ and $M_2$ can be glued together along a shared boundary $\Sigma \subseteq \partial M_1, \partial M_2$ by using the symmetric bilinear $\langle \Phi^*, \Psi \rangle_\Sigma$ to evaluate the pair of states at $\Sigma$.  In other words, one requires the fields at the boundary to match, and integrates over all possible fields while being careful to mod out by the newly introduced gauge symmetries.  By doing this, one obtains $Z$ for the combined region $M_1 \sqcup_\Sigma M_2$, where $\sqcup_X$ denotes the union of two disjoint submanifolds, glued together at a shared boundary $X$.\footnote{This subscript will sometimes be dropped when the identity of $X$ is unimportant to the discussion, but the step where the manifolds are glued together at their shared boundary is still implied.}

\medskip\textbf{Superselection Sectors.}  Suppose that a spatial slice $\Sigma$ is decomposed into two open regions joined by a boundary $B$, so that $\Sigma = \Sigma_1 \sqcup_B \Sigma_2$.  We would like to be able to say that the Hilbert space $\mathcal{H}_\Sigma$ decomposes into tensor factors $\mathcal{H}_{\Sigma_1} \otimes \mathcal{H}_{\Sigma_2}$.  However, this is not quite true since the normal components of the electric and magnetic fields must match across the boundary $B$.  Since these fields can be measured on both $\Sigma_1$ and $\Sigma_2$, by microcausality they must commute with all other operators localized in either $\Sigma_1$ or $\Sigma_2$, i.e. from the perspective of a single region they are superselection sectors.

In order to implement the magnetic constraint, we require that 1) the choice of bundle on $B$ must match on both sides, and 2) so must the connection; in other words the vector potential $A_B$ must agree up to gauge transformations (including large ones).  Note that if $B$ has noncontractable curves, this constraint also requires the holonomy of $A$ around those curves to agree.

The electric constraint has to do with $G_B$, the group of gauge transformations on $B$.  The fields in a single region (e.g. $\Sigma_1$) do not have to be invariant under gauge transformations which affect the boundary.  Instead they can transform under nontrivial unitary representations of $G_B$.  Since $G_B$ is abelian, the irreps are all 1-dimensional, and are labelled by the choice of electric flux across $B$.  However, the state of $\Sigma$ must be invariant under $G_B$, so the irrep of $\Sigma_1$ must be conjugate to the irrep of $\Sigma_2$.  This enforces the constraint that the normal component of the electric field matches across the boundary.

\medskip\textbf{Evolution Operator.}  Now suppose we divide the boundary of a Euclidean spacetime region into two parts joined by $B$, so that $\partial M = \Sigma_1 \sqcup_B \Sigma_2$.  Then the partition function can also be used to calculate a time evolution operator $\mathcal{E}_\mathrm{cov}(\Sigma_1 \to \Sigma_2)$ which evolves a state of the fields in $\Sigma_1$, to a state of the fields in the complement $\Sigma_2$, while preserving superselection data at $B$:
\begin{equation}
\mathcal{E}_\mathrm{cov} = (gJ \otimes I) Z(M),
\end{equation}
where $J$ is the antiunitary time-reversal operator (i.e. complex conjugation of the position-space wavefunction), and $g$ is the inner product defined in Eq. (\ref{inner}).  Here $Z(M)$ is regarded as a state using Eq. (\ref{state}), $gJ$ is an operator whose domain is the Hilbert space $\mathcal{H}_{\Sigma_1}$ but whose target is the dual vector space $\mathcal{H}^*_{\Sigma_1}$, and $I$ is the identity acting on $\mathcal{H}_{\Sigma_2}$.  Thus $\mathcal{E}_\mathrm{cov}$ takes states from  $\mathcal{H}_{\Sigma_1}$ to $\mathcal{H}_{\Sigma_2}$.

\medskip\textbf{Wick Rotation.}  We now wish to define $\mathcal{E}_\mathrm{cov}$ for a globally hyperbolic Lorentzian manifold $M$ with boundary $\partial M = \Sigma_1 \sqcup_B \Sigma_2$, such that $\Sigma_1$ is an initial spacelike slice and $\Sigma_2$ is a final spacelike slice.  To do this, we consider the class of all Euclidean metrics for which $\partial M$ has the same geometry.  $\mathcal{E}_\mathrm{cov}$ is defined in terms of the partition function $Z(M)$, which in turn depends  on the metric $g_{\mu\nu}$.  We will assume without proof that, as usual in QFT, $Z(M)$ is an analytic function of $g_{\mu\nu}$, which can be analytically continued through complex values of $g_{\mu\nu}$ to the Lorentzian signature, without encountering obstructions or ambiguities.  Because the Euclidean partition function is real, the Lorentzian partition function is invariant under time-reversal $T$.

Converting from $Z(M)$ to $\mathcal{E}_\mathrm{cov}$ does not depend on the signature, and can therefore be done either before or after the Wick rotation.

\medskip\textbf{Canonical Approach.}  Given a particular folation of a globally hyperbolic Lorentzian manifold (with boundary) into Cauchy surfaces $\Sigma(t)$, we will assume that there is some way to define a canonical theory of free electromagnetic fields, in terms of the on-shell degrees of freedom on each time $\Sigma(t)$, modulo gauge transformations.  Rather than constructing this theory, we will take a more axiomatic approach and list the properties which we expect it to have.

In the canonical approach, the evolution operator $\mathcal{E}_\mathrm{can}(\Sigma_1 \to \Sigma_2)$ must be a manifestly unitary evolution operator.  (On a non-static background, the evolution operator $\mathcal{E}$ is actually a map between two \emph{different} Hilbert spaces.  When we call such an evolution operator ``unitary'', we really mean that it is an isomorphism between these two Hilbert spaces.)  However, we do not assume that $\mathcal{E}_\mathrm{can} = \mathcal{E}_\mathrm{cov}$, or even that $\mathcal{E}_\mathrm{can}$ is necessarily covariant under a change of foliation.

Instead, we assume that the canonical and covariant theories both satisfy the classical linear field equations, interpreted as operator equations.  Using these equations of motion, we can fix the evolution of the wavefunction from $\Sigma_1$ to $\Sigma_2$ up to an overall numerical factor $X$ which can depend only on the geometry of $M$:
\begin{equation}\label{X}
 Z_\mathrm{cov}(M, A_{\partial M}) = Z_\mathrm{can}(M, A_{\partial M}) X(M).
\end{equation}

\subsection{Argument for unitarity}

We will now argue that $\mathcal{E}_\mathrm{cov}$ is unitary in Lorentz signature.  The normal method for proving unitarity is to show that the covariant theory is equivalent to the canonical formulation using a reduced phase space.  However, according to Ref.~\cite{Vassilevich1994} these two formulations are not in general equivalent.  Nevertheless, it is still extremely useful to know that a manifestly unitary canonical formulation of Maxwell exists, because we can use Eq. (\ref{X}) to rewrite the covariant evolution operator as
\begin{equation}
\mathcal{E}_\mathrm{cov} = X \mathcal{E}_\mathrm{can} = |X|U,
\end{equation}
where $X$ is a complex number, and $U$ is some unitary evolution rule (i.e. an isometry).  In the second equality we have performed a polar decomposition of $X$, and absorbed the phase into $U$.  After properly renormalizing ultraviolet divergences, both evolution operators should be finite, so $X \ne 0$.  Hence one can define a unitary evolution operator as:
\begin{equation}\label{Ucov}
\frac{\mathcal{E}_\mathrm{cov}}{|X|} = U.
\end{equation}
Although we have used $\mathcal{E}_\mathrm{can}$ to prove the existence of $U$, once we know it exists it can be defined using $\mathcal{E}_\mathrm{cov}$ alone, by choosing the one and only positive real number $|X|$ that makes the right-hand-side unitary.  So by Curie's principle (unique solutions to symmetric problems have symmetric solutions), $U$ and $|X|$ are themselves covariant and T-invariant.

We will now show that $|X| = 1$, i.e. $\mathcal{E}_\mathrm{cov}$ is unitary.  First we observe that $|X|$ by construction depends only on the geometry, not the state.  Next we exploit the gluing property of Eq. (\ref{unfix}):
\begin{equation}
\mathcal{E}_\mathrm{cov}(M_1 \sqcup M_2) = \mathcal{E}_\mathrm{cov}(M_1) \mathcal{E}_\mathrm{cov}(M_2).
\end{equation}
Because the product $U(M_1)U(M_2)$ of two unitary operators is unitary,
and $U(M_1 \sqcup M_2)$ is the unique unitary operator equal to a real number times $E_\mathrm{cov}(M_1 \sqcup M_2)$,
it follows that
\begin{equation}
U(M_1 \sqcup M_2) = U(M_1) U(M_2).
\end{equation}
Hence $|X|$ is also local:
\begin{equation}
|X|(M_1 \sqcup M_2) = |X|(M_1) |X|(M_2).
\end{equation}
Using these relations, we can chop up the manifold into tiny pieces, in order to show that $|X|$ is given by a local action:
\begin{equation}
|X| = e^{iS[M]}.
\end{equation}
Here the action $S$ is imaginary, and depends only on the metric $g_{ab}$.  This action might include boundary terms at $\partial M = \Sigma_1 + \Sigma_2$.  Temporarily ignoring these boundary terms, we can expand out the action in derivatives of the metric to obtain
\begin{equation}\label{iL}
S = i \int_M d^Dx \sqrt{-g} [\,a + b R + (c R^2 + d R_{ab}^2 + e R_{abcd}^2) + \ldots\,].
\end{equation}
When this action is Wick rotated back to the Euclidean signature, it remains purely imaginary.  This is because each power of the Riemann tensor has an even number of time-derivatives, while the $i$ coming from rotating the volume form $\sqrt{-g} = i\sqrt{g}$ turns $e^{iS}$ into $e^{-S}$.  It follows that such terms cannot exist by $T$-invariance.  (Although this fact could be deduced from the Lorentz signature theory, it is easiest to diagnose violations of $T$ in Euclidean signature since there $T$ consists of complex conjugation only.)  Therefore the covariant partition function is unitary up to a local boundary action.

Terms in the boundary action could depend on the metric on the boundary, as well as its normal derivatives.  However, it turns out that no boundary term can be found having the necessary properties.  We start by mapping each of the initial and final Hilbert spaces into a specific Hilbert space $\mathcal{H}$ described by an orthonormal basis of states $(\psi_1, \psi_2, \psi_3 ...)$.  This allows $U$ to be regarded as a unitary operator in $\mathcal{H}$.  In this basis, the inner product $\langle \Phi, \Psi \rangle$ can be written as the identity matrix.  Since raising and lowering with respect to the identity matrix is trivial, the boundary terms must take the same form for the ``initial'' slice $\Sigma_1$ and the ``final'' slice $\Sigma_2$.

The absence of allowed boundary terms is easiest to see if $\ln |X|$ is rotated back to Euclidean signature.  In order to satisfy the gluing property, the boundary term would have to contain an odd number of normal derivatives.  (Terms such as $\mathrm{Vol}(\partial M)$ containing an even number of derivatives would not cancel when gluing together two regions.)  T-reversal symmetry says the boundary term has to be real.  Now an \emph{odd, real} boundary term in Euclidean signature Wick rotates into a contribution to the Lorentzian $\ln |X|$ which is \emph{odd} and \emph{imaginary}.  But $\ln |X| = iS$ is real in Lorentz signature by construction (and in any case, an imaginary contribution to $iS$ would not lead to a violation of unitarity.)

Since no contributions to $|X|$ are possible, it follows that $\mathcal{E}_\mathrm{cov}$ is unitary.  This is true even if it disagrees with $\mathcal{E}_\mathrm{can}$ on nonstatic manifolds.

\section{Discussion}

In section \ref{2}, we defined the covariant electromagnetic partition function on spatially compact manifolds, taking into account issues involving zero modes of the field and gauge symmetry, as well as nontrivial U(1) bundles and large U(1) gauge transformations.  

Section \ref{3} showed how to implement the Faddeev-Popov trick on compact Euclidean manifolds.  In Eq. (\ref{fixed}) we found several kinds of departures from the usual ``vector minus two scalars'' partition function.  In addition to summing over nontrivial bundles, modding out by large gauge transfromations (i.e. Gribov copies), and omitting ghost zero modes, there is also a nontrivial normalization factor $(q / \mu_\alpha^2) \sqrt{\xi / 2\pi V}$, depending on the charge $q$, the t'Hooft gauge parameter $\xi$, the spacetime volume $V$, and the dimensionful parameter $\mu_\alpha$ which defines the measure on the space of gauge-transformations.  

Because the normalization factor depends on the geometry through $V$, it contributes to the vacuum stress-energy tensor, and to the entropy of e.g. de Sitter space.  In electromagnetism with one spacetime dimension (which has no degrees of freedom at all), this factor is necessary to obtain a trivial partition function.

We also showed that the covariant approach is local and unitary, although the Faddeev-Popov trick obscures these properties.  In our proof of unitarity (section \ref{unitary}), we relied on the existence of a canonical formulation of electromagnetism, but we did not explicitly construct the canonical formulation, nor did we prove that the two approaches are generally equivalent.

We were able to explicitly compare to the canonical formulation for two special categories of Euclidean manifolds (section \ref{calc}).  One was $D = 2$ electromagnetism (where there are no local degrees of freedom), treated in \cite{Cordes1994,Donnelly2012}.  The second was ultrastatic manifolds of the form $S^1 \times \Sigma$, where the canonical partition function takes the form $\mathrm{tr}\,e^{-\beta H}$ summed over physical modes. In both of these cases, we found an agreement with the covariant approach.  Although we have not done the calculation, we expect that the two approaches will always agree for static, horizonless manifolds.

However, we expect that serious complications will arise for $D > 2$ time-dependent manifolds.  In these cases, the two ghosts do not in general cancel out the temporal and longitudinal modes of $A_\mu$ \cite{Vassilevich1994,Donnelly2012}.  In other words, the unphysical modes give rise to physical effects, which are hard to explain in terms of the canonical degrees of freedom.  Similarly, in the case of manifolds with horizons, there are effects arising from the fact that vectors and scalars have different boundary conditions at the horizon.  This seems to be related to the mysterious ``contact term'' discovered by Kabat \cite{Kabat1994}, who calculated that Maxwell fields contribute a negative divergence to the horizon entropy.  This contact term cannot come from the entanglement entropy of the canonical modes, which is intrinsically positive.  A similar discrepancy appears for gravitons \cite{Fursaev1996,Solodukhin2011}.

In fact, the original motivation of this work was to investigate the validity of this contact term on compact manifolds (where one need not worry about complications due to a boundary at infinity).  In an earlier article \cite{Donnelly2012}, we showed that the contact term is absent in $D = 2$ for canonically quantized electromagnetism.  This article confirms that the covariant $D = 2$ theory is equivalent, so the contact term is absent there as well.  The case of higher dimensions will be treated in a future article.

Nevertheless, quite aside from the contact term, it would be disturbing if the covariant and canonical formulations gave different answers to physical questions.  This article goes halfway towards resolving the issue, by showing how to properly treat the covariant formulation.  Further progress on the discrepancy will require a better understanding of the canonical approach.

It is not surprising that little work has been done on canonical Maxwell fields on nonstatic spacetimes, because the required conceptual apparatus is technically difficult.  For each time slice, one has to construct a Hilbert space based on physical modes only.  In order to write down the dynamics in terms of a time-dependent Hamiltonian, one must first select an isomorphism between the Hilbert spaces at each moment of time (on a static manifold one can use time translation symmetry to do this, but on a nonstatic manifold one has to make some arbitrary choices).  At the end, one has to prove covariance, i.e. that the dynamics would have been the same regardless of the choice of slicing.

In Vassilevich's approach to canonical QFT, one slices up the spacetime, identifies the ``physical modes'' which would contribute to the canonical analysis, and then quantizes these modes using covariant methods \cite{Vassilevich1994}.  Vassilevich found a noncovariant result, which disagrees with the covariant approach for e.g. Euclidean de Sitter space $S^4$ or $S^2$.  However, Ref.~\cite{Vassilevich1994} did not include the above normalization factor in his canonical approach.  It also used the wrong measure in the path integral.  (In the $D = 2$ case, the path integral measure on an expanding spacetime is a function of time.  The correct measure is easiest to calculate if one uses the integrated spacetime volume as the  ``time'' parameter, as in Ref. \cite{Donnelly2012}.)  It may be that a more sophisticated version of the canonical approach might give an answer which agrees with the covariant approach.

The analysis of section \ref{unitary} shows that the canonical and covariant approaches agree up to a geometry-dependent phase factor.  Therefore, so long as one is willing to insert such phases as a ``fudge factor'', the canonical method can be made to work.  The question remains whether these phases can be derived elegantly in a canonical framework.

In Wald's approach to canonical QFT in curved spacetime \cite{Wald1994}, one identifies the physical modes of the field in a slice-independent way, and quantizes this using the symplectic structure.  This approach is manifestly covariant, but Wald was not able to directly calculate the stress-energy tensor.  Instead he lists some axioms (including covariance) which the stress-energy tensor should satisfy, and used this to pin down their correct form.

Assuming this approach can be applied to gauge fields (which Ref. \cite{Wald1994} claims, but does not work out in detail), it should be possible to derive the evolution operator $\mathcal{E}_\mathrm{can}$, use this to prove that $\mathcal{E}_\mathrm{cov}$ is unitary, and then show that the covariant stress-energy tensor obeys Wald's axioms.  Unfortunately, this roundabout method obscures the physical origin of the effects.  It would be good to have a more direct way to derive the stress-energy tensor from the canonical formulation.  If our treatment of the covariant calculation is any indication, this will probably require being very careful when handling the gauge symmetries.

\section*{Acknowledgments}
We are grateful for discussion with Ted Jacobson, Dan Kabat, Sergey Solodukhin, Joe Polchinski, Don Marolf, Bernard de Wit, Amanda Peet, Ashoke Sen, and David Berenstein.
A.W. is supported by the Simons Foundation.

\bibliographystyle{utphys}
\bibliography{compact}

\end{document}